\documentstyle[a4,12pt]{article}

\begin{document}

\def\ben{\begin{equation}}
\def\een{\end{equation}}
\def\half{{1 \over 2}}
\def\pr{{\em Phys. Rev.~}}
\def\bea{\begin{eqnarray}}
\def\eea{\end{eqnarray}}

\rightline{\small{KEK-TH/472}}
\rightline{\small{TIFR-TH/96-11}}
\vskip 1.5cm

\begin{center}
{\large{ \bf BLACK HOLE ENTROPY AND STRING THEORY }}
\end{center}
\vskip 0.5cm
\begin{center}
Sumit R. Das 
\vskip 0.4cm 
{\em Theory Group, KEK} \\
{\em Tsukuba-shi, Ibaraki-ken 305, JAPAN
\footnote{Address till April 12, 1996}}\\
and \\
{\em Tata Institute of Fundamental Research} \\
{\em Homi Bhabha Road, Bombay 400005, INDIA
\footnote{Permanent Address}}

\end{center}

\vskip 1.5cm
\begin{center}
{\bf Abstract}
\end{center}

{\footnotesize 

\parbox{16cm}{This is an expanded version of a talk given at ``{\em
IInd Recontre du Vietnam}'' held at Ho Chi Minh City in October, 1995.  We
discuss several aspects of black hole entropy in string theory. We
first explain why the geometric entropy in two dimensional noncritical
string theory is nonperturbatively finite. We then explain the
philosophy of regarding massive string states as black branes and how
the Beckenstein-Hawking entropy for extremal BPS black holes may be
understood as coming from degeneracy of string states.  This is then
discussed in the context of D-strings in Type IIB superstrings.  We
then describe non-BPS excitations of D-strings and their entropy and
explore the possibility that their decay describes Hawking radiation.
For these D-strings and other D-branes the entropy and temperature are
consequences of the physical motion of stuck open strings along the
D-brane and this leads to a simple space-time interpretation. Finally
we speculate that the horizon may be itself regarded as a D-brane.}}
\vskip 1.0cm

In this contribution I will discuss some aspects of black hole entropy and
how string theory has helped us to understand this rather
mysterious quantity. This is based on a talk given at {\em IInd Recontre
du Vietnam}. However I have added some developments which
took place very recently to make the article meaningful at this time.

Even before the phenomenon of black hole radiation was discovered,
Beckenstein \cite{beck} noticed a profound analogy between black holes
in classical general relativity and the laws of thermodynamics.  In
particular he found that the surface gravity at the horizon of a black
hole can be regarded as a temperature and the area of the horizon as
an entropy. Hawking \cite{hawking} showed that black holes radiate due
to quantum effects with a temperature which is indeed proportional to the
surface gravity and this gave a precise formula for the ``Beckenstein-
Hawking entropy'' $S_{HB}$ which, for simplest black holes, read
\ben
S_{HB} = {A \over 4 G_N}
\label{eq:one}
\een
where $A$ is the horizon area and $G_N$ is Newton's constant. In the
presence of matter or higher derivative terms in the action this
formula may require modification \cite{visser}.

There is a way in which the answer (\ref{eq:one}) follows from a
{\em classical} calculation in euclidean gravity \cite{gibbhawk}. One
computes the action of the euclidean black hole solution and
identifies it with a free energy. Standard thermodynamics together
with the formula for the temperature gives rise to the expression
(\ref{eq:one}).

It has always been a big mystery why this is an entropy, since there is
no obvious statistical origin of this quantity. As we will discuss later,
string theory is offering a way to resolve this mystery.

\section{Geometric Entropy and 2d strings}

Before delving into this, let us discuss another related object which
enters in black hole thermodynamics. This is ``entanglement entropy''
\cite{gent} \cite{suss1} \cite{suss2}
or ``geometric entropy'' of black holes and measures, in a given
quantum state, the correlation of the region inside the horizon with
the region outside. In usual field theories of scalars and spin-$1/2$
particles, this is the same as the quantum correction to the Hawking-
Beckenstein entropy defined through the partition function on the
eulidean background with a defect angle. For gauge fields the two
quantities are not the same \cite{kabat}. In string theory the
situation is tricky since a string can be partly inside and partly
outside the horizon \cite{suss1} \cite{suss2}. 
We will come back to this aspect in a later
section.

\subsection{Geometric entropy}

The notion of a geometric entropy exists in any theory, regardless of
black holes. Consider for example some field theory in $d + 1$
dimensions.  Let us divide the space into two halves by putting in an
imaginary plane at $x = 0$ where $x$ is one of the spatial
coordinates. The wave functional in any given state may be then
written as $\Psi [\phi_L, \phi_R]$ where $\phi_L$ and $\phi_R$ are the
fields in the regions $x < 0$ and $x > 0$ respectively. 
If we are interested only in operators which depend on $\phi_R$ all
observables may be expressed in terms of a reduced density matrix
$\rho[\phi_R, \phi_R']$
\ben
\rho[\phi_R, \phi_R '] = \int {\cal D}\phi_L~
\Psi [\phi_L, \phi_R] \Psi^* [\phi_L, \phi_R ']
\label{eq:two}
\een
One can now associate an entropy $S_G$ with this density matrix 
\ben
S_G = - {\hat \rho} {\rm log}~{\hat \rho}
\label{eq:three}
\een
where ${\hat \rho}$ is the normalized density matrix.

In usual flat space the divide at $x = 0$ is imaginary.  However for
an external observer in a black hole background, the horizon provides
a real division of space.  Thus if we define an object like
(\ref{eq:two}) with $L$ and $R$ being replaced by the interior and the
exterior of the horizon respectively we have a physical quanity which
measures the amount of information hidden inside the horizon. In a
dynamical situation of collapse and subsequent evolution one may then
measure the net change in the total entropy $S_{HB} + S_G$ and see
whether this increases. This is indeed possible
\cite{twodent1},\cite{twodent2} in CGHS-RST models of
black hole evolution.

An important property of the entanglement entropy is that in field
theory this quantity depends on the ultraviolet cutoff
\cite{thoft}.  In
$d+1$ dimensions it behaves as $\Lambda^{d-1}$ for $d > 1$ and ${\rm
log}~(L\Lambda)$ for $d = 1$ where $\Lambda$ is a (momentum)
ultraviolet cutoff and $L$ is the size of the system. 
In a curved background one must use
an invariant cutoff and as shown e.g. in \cite{twodent2} this means
that the entanglement entropy depends on the metric at the horizon so
that the time evolution of the latter determines the time evolution of
the entropy. For the two dimensional models this implies that an
indefinite amount of information is lost - so long as we trust the
semiclassical approximation in regions of low curvature
\cite{twodent2}.

There is a simple way to understand why the geometric entropy is
logarthimlically divergent in $1 + 1$ dimensional field theories
\cite{twodent2}. Consider
for example a free massless scalar field $\phi(x,t)$ in a spatial box
of size $L$. We want to find the entropy of entanglement between the 
left and the right halves of a divide at $x = 0$. Now write the field
in a basis of non-overlapping wave packets formed out of wave numbers
$ 2^j k_0 < k < 2^{j+1} k_0 $ where $k_0 = {2\pi \over L}$ and $j$ is
some integer. For each value of $j$ only one wavepacket straddles
across $x = 0$ and contributes to the entropy and the contribution
is independent of $j$ due to conformal invariance. Thus the entropy
is proportional to the maximum allowed value of $j$, $j_{max} \sim {\rm log}
(L\Lambda)$. A similar analysis gives the entropy for entanglement between
some region $x_1 < x < x_2$ and the rest of space. Here the relevant 
contribution comes from wavenumbers $k > 1/(x_2 - x_1)$ and one has
\ben
S_G = {\rm log}~[(x_2-x_1)\Lambda]
\label{eq:ffour}
\een
The entropy is infinite simply because there are an infinite number
of modes which contribute to it.

In perturbative string theory there are no ultraviolet divergences,
hwoever 
there is now an infrared divergence associated with a
hagedorn transition near the horizon \cite{barbon}. The fact
that the high temperature behavioor of string theory enters into the
discussion implies that the issue is nonperturbative.

We know very little about nonperturbative behavior of critical string
theories, though it is certainly hoped that the recent developments in
string duality will teach us something. We do have a nonperturbative
formulation for at least one noncritical string theory - the two dimensional
string defined through the double scaling limit of matrix quantum mechnaics
\cite{deqone}. However we dont understand black holes in the matrix 
model formulation. Nevertheless, as we just saw, the ultraviolet divergence
of the geometric entropy of field theories is present in usual flat space
and one may ask the same question in this two dimensional string theory.
In the following we investigate this question. Details may be found
in \cite{dasent}.

\subsection{Geometric entropy in the 2d string}

The fundamental formulation of this theory is in terms of mutually
noninteracting nonrelativistic fermions in an inverted harmonic
oscillator potential $V(x) = - x^2/2$. 
\ben
S = \half \int dx dt ~\psi^\dagger (x,t)
[i\partial_t + \partial_x^2 + x^2]
\psi (x,t)
\label{eq:threea}
\een
Let the fermi level be denoted by
$-\kappa$. Then the excitations of the theory are described in terms of a
massless scalar field $\xi (x,t)$ which is the fluctuation of the
density of these fermions. This massless scalar field is related to
the only poropagating degree of freedom of two dimensional string
theory - the massless ``tachyon'' - but not the same \footnote{See
e.g. Polchinski in \cite{deqone}}. The dynamics of $\xi (x,t)$ is
described in perturbation theory by the collective field hamiltonian
\ben
H = \int d\tau [\half (\Pi_\xi^2 + (\partial_\tau \xi)^2)
+ {1 \over 6 \rho_0^2 (x)}((\partial_\tau \xi)^3 +
3 \Pi_\xi (\partial_\tau \xi) \Pi_\xi)]
\label{eq:four}
\een
where
\ben
\rho_0^2 (x) = x^2 - \kappa = {1 \over g_{eff}(x)}
\label{eq:five}
\een 
is the inverse (position dependent) string coupling in the theory,
and $\tau$ is a new spatial coordinate defined by $dx = \rho_0(x) d\tau$
The theory is thus strongly coupled near
the hump of the potential at $x = 0$ and weakly coupled in the
asymptotic region $ |x| >> {\sqrt{\kappa}} $. In terms of $\tau$
one has $g_{eff}^{-1} (\tau) = \kappa {\rm sinh}^2~\tau$ so that
for a given $\tau$ large $\kappa$ means weak coupling.

In a sense (\ref{eq:four}) is a ``string field theory'' whose fields
are directly related to the perturbative excitations of the model.
The position dependence of the string coupling is typical of noncritical
strings and comes from a nontrivial dilaton background.
On the other hand (\ref{eq:threea}) is a more basic description of
the theory.

In perturbation theory one has a massless scalar field and the lowest
order answer would have the
characteristic logarithmic dependence on the cutoff.  Note that this
is not the standard ultraviolet divergence of vacuum loop diagrams :
in the collective field theory such diagrams are finite as expected in
string theory. As explained in \cite{dasent} this is in fact related
to the high temperature behaviour of the theory.  In principle one
could take into account the effect of interactions using the
hamiltonian (\ref{eq:four}) : this would be like computing the
geometric entropy in string perturbation theory and it would be hard
to imagine how this dependence on the cutoff is removed.

However here we are lucky to have a formulation of the model which
is as easily analyzable in strong coupling as in weak coupling - viz.
the fundamental fermionic formulation given by (\ref{eq:threea}). 

In fact the essential physics is present even for {\em free}
nonrelativistic fermions in a box which
 we discuss first. Let $k_F$ denote the
fermi momentum so that all states in $-k_F < k < k_F$ are filled in
the ground state. In the following we will define a shifted momentum
$q = k - k_F$ with corresponding redefined fields (by a phase) $\chi
(q) = \psi (k_F + q)$.  The ground state wave functional may be
written in terms of the grassmann valued fields $\chi (q)$ as
\ben
\Psi_0 = {\rm exp}[-\half \int_{-\infty}^{\infty} dq~ {\bar \chi}(q)
\chi (q) + \int_{-2k_F}^0 dq~{\bar \chi}(q) \chi (q)]
\label{eq:six}
\een
In terms of position space fields the first term in the exponent is 
an integral of a {\em local} density. This does not entangle the
left and right halves and does not contribute to the entropy. The entire
contribution comes from the second term which involves only modes in
the fermi sea. 

This is the crucial point. Since only modes in the fermi sea
contribute and the fermi sea has a finite depth, 
the ultraviolet cutoff does not play a role in the
geomteric entropy - as opposed to the case of the scalar field
discussed above. However, if we expand around large $k_F$, in the
lowest order the fermi sea of nonrelativistic fermions becomes the
infinite Dirac sea. One effectively has relativistic fermions which is
equivalent in two dimensions to a relativistic boson. In this limit
one gets the usual logarithmically divergent answer.

The bosonized form of the theory of free nonrelativistic fermions is
given by the collective field theory (\ref{eq:four}) with $\rho_0$ in
(\ref{eq:five}) replaced by $\rho_0^2 = \mu_F = \half k_F^2$. Thus
exapnsion around $k_F = \infty$ is the perturbation expansion of
the bosonic theory. We immediately see that the finiteness of the
geometric entropy is an essentially {\em nonperturbative} phenomenon in
the bosonic description.

Let us now go back to the one dimensional string, i.e. include the
effect of a $-x^2$ potential. 
The problem we want to consider is the following : consider a box of
size $l$ centered around a point $x = - x_0$. Choose $x_0$ to lie in the
asymptotic region where the theory is weakly coupled and $l$ to be much
smaller than the overall size of the system. We want to estimate the
entanglement entropy between this box and the rest of the system.
The depth of the fermi sea at $x_0$ is (in energy) $x_0^2 - \mu$. Thus
in this calculation the relevant range of wavelengths is 
${\sqrt{x_0^2 - \kappa}} < \lambda < l$. The answer for the
geometric entropy becomes \cite{dmatone}
\ben
S_G \sim {\rm log}[l^2 g_{eff}(x_0)]
\label{eq:seven}
\een
where we have used (\ref{eq:five}). The answer does not depend
on the ultraviolet cutoff and is clearly nonperturbative.

The above calculation drives home a point which may be quite generic in
string theory : the true degrees of freedom at high energies may be
rather different and fewer in number than what one expects from
perturbative considerations. For critical string theory we dont know
what these degrees of freedom are. For the two dimensional string the
degrees of freedom are transparent since we have free nonrelativistic
fermions.

\section{The Beckenstein Hawking Entropy in Superstring Theory}

We now turn to the question of Beckenstein-Hawking entropy.  It is an
old idea that very massive elementary particles behave as black holes
in strong coupling\cite{salam}. Recently this idea has proved to be
very fruitful in understanding the Hawking-Beckenstein entropy in
string theory. The idea is as follows. It is well known that in string
theory there are very massive states and there are large number of
states of a given mass. Could it be that this degeneracy of states is
the origin of the Beckenstein-Hawking entropy \cite{nano1}
\cite{suss1}
\cite{suss3} ?

However at the face of it, the idea seems to run into trouble for usual
Schwarzschild black holes. Here the entropy $S$ for mass $M$ is
$S \sim M^2$ whereas it is known that in string theory the degeneracy
of states grows as $e^{M}$. It was argued in \cite{suss1}, \cite{suss3} that 
quantum effects might renormalize the mass suitably. A different proposal
involving noncritical strings has been put forward in
\cite{nano2}

Luckily there are states in string theory whose masses are not
renormalized - these are the BPS staurated states. The above idea
may be tested for such states which are to be identified with extremal
black holes \cite{vafsuss}. Indeed 
such states do behave like extremal black holes in scattering processes
\cite{duff}\cite{callanpeet}\cite{mandalwadia}. However, most of
these extremal holes have zero horizon area and should lead to a zero
entropy ! In \cite{sen} it was, however, proposed that the area of the
{\em streched horizon} \cite{suss2} rather than the event horizon
should be identified with the entropy. Indeed for a class of such BPS
saturated extremal holes in heterotic string theory compactified on
$T^6$ it was found that the dependence of the streched horizon area on
the parameters sepcifying the solution agrees with the dependence of
the degeneracy of string states on these parameters (which are
generically charges) \cite{sen} \footnote{
It has also been argued using thermodynamic
arguments that the entropy of an extremal black hole {\em could} be proportional to the mass \cite{pmitra}}.

\subsection{D-branes}

Recently this connection has been better understood owing to
the realization that a certain class of solitons in superstring
theories can be described in terms of objects called D-branes
\cite{dbranes}. 

Consider a theory of closed strings, e.g. the Type IIB superstring
theory.  Now add to this some open strings whose ends are
restricted to move only in $p$ of the spatial
dimensions. This means we have imposed Dirichelt boundary conditions
on $(9-p)$ coordinates. The values of these $(9-p)$ coordinates then
define a $p$- dimensional object moving in time - a $p$-brane. This
$p$ brane is a soliton in the closed string theory. The collective
coordinates which describe the low energy excitations of this soliton
are precisely the lowest mass modes of the open strings whose ends are
stuck on the brane.

Closed superstring theories have generally two types of gauge fields
coming from the NS-NS and R-R sectors on the world sheet. Consider
for example the Type IIB theory. The ten dimensional gauge fields of
this theory consist of (1) NS-NS sector : dilaton $\phi$, metric
$g_{AB}$ and antisymmetric tensor field $B^{(1)}_{AB}$ (2) R-R sector :
an axion field $\chi$, antisymmetric tensor $B^{(2)}_{AB}$ and a
rank four gauge field with self dual field strength $C_{ABCD}$.
The object which carries charges under the NS-NS
gauge field $B^{(1)}_{AB}$ is in fact the elementary string itself.
At low energies the elementary string is described by a classical
solution of the effective field theory - the macroscopic string
\cite{dharvey}.  However there are no states in the perturbative
spectrum of the elementary string theory which carry charges of the
R-R gauge field $B^{(2)}_{AB}$.  Remarkably, these missing states
carrying R-R charges are D-branes \cite{polchin}.  The field content
shows that these D-branes describe $p$-branes with odd $p$ for Type IIB
and even $p$ for Type IIA.

The Type-IIB theory is conjectured to be self-dual under an $SL(2,Z)$
symmetry which transform $(\phi, \chi)$ and
$(B^{(1)}_{AB},B^{(1)}_{AB})$ into each other. In fact there are an
infinite number of classical string solutions labelled by $(m,n)$
where $m$ denotes a quantized NS-NS charge and $n$ a quantized R-R
charge \cite{schwarz}. These solutions are in fact low energy
descriptions of bound states of D-branes and
elementary strings \cite{witten}.  In particular the string with
$(0,1)$ charge has a nonzero $B^{(2)}$ (plus metric and dilaton) but
zero $B^{(1)}$ - we will call this the D-string.  The duality
conjecture then implies that the states of the D-string behave exactly
like the states of the elementary string with the coupling replaced by
the inverse coupling.  Evidence for this has accumulated over the past
few months \cite{witten} \cite{senvafa}.

\subsection{BPS States and solutions}

Consider flat ten dimensional space with one of the directions, say
$X^9 = z$ compactified on a circle of circumference $L$. Then the
mass of an elementary string state is given by
\ben
M^2 =  (n_w L T + {2\pi n_p \over L})^2 + 8\pi T N_R 
 =  (n_w L T - {2\pi n_p \over L})^2 + 8\pi T N_L
\label{eq:nine}
\een
where $n_p$ is the quantized momentum in the $X^9$ direction and $N_L,
N_R$ denotes the oscillator number of the left and right moving
oscillators on the world sheet. We have defined them so that the
minimum values ($1/2$ for NS and $0$ for R sectors) have been
subtrated out. $T$ is the elementary string tension.  The mass formula
(\ref{eq:nine}) is perturbative. However there are a
special class of states when this is in fact exact. These are the BPS
saturated states which have {\em either} $N_R = 0$ with $N_L$
arbitrary {\em} or $N_L = 0$ with $N_R$ arbitrary. 
Consider the first case, $N_R = 0$. The level matching condition
(\ref{eq:nine}) then shows that $N_L = 4 n_w n_p$. 

Thus for a given
$(n_p, n_w)$ there are many states in the string theory : these correspond
to the number of ways one can have a total level $N_L$ from the basic
bosonic and fermionic oscillators. As is well known the number of such
states grows as $\sim e^{2\pi{\sqrt{2N_L}}}$ for large $N_L$.
The low energy description of such states are oscillating macroscopic
strings with string tension $\sigma = n_w T$ and a
momentum density $p = 2\pi n_p / L^2$ along the string
\cite{callanpeet} \cite{dabgaunt}
and carry the NS-NS charge for the gauge field
$B^{(1)}$.  These are in fact extremal black strings with the horizon
coinciding with the singularity. If five of the transverse dimensions
as well as the string direction is compactified they appear as
charged black holes in four dimensions carrying two kinds of $U(1)$ charge :
one coming from the $B^{(1)}$ charge and one from the $g_{z\mu}$ component
of the metric. This is in fact one of the solutions considered in
\cite{sen} though in that case the origin of the charges is rather
different.

The classical solution representing this black string (or black hole
in the compactified theory) is specified by only the string tension and the
charge density. However identification of this with a BPS string state shows
that there are many possible states with the same set of charges, 
corresponding to the many ways of making $N_L$ from the basic string
oscillators. It is therefore, logical to assign an {\em entropy} to such a
black string which is the logarithm of the degeneracy. For large charges
\ben
S = 2\pi {\sqrt{8 n_w n_p}} = L {\sqrt{2\pi \sigma p}}
\label{eq:ten}
\een
As mentioned earlier this agrees with the area of the streched horizon.

We now describe the excitations for the D-string carrying RR charge
rather than NS-NS charge \cite{dasmath}. The classical solutions may
be obtained by the duality transformation described above. The
solution is the same as the oscillating string NS-NS solution with
the following changes : (1) the elementary string tension $T$ is
replaced by the D-string tension $T_D = T/g$ where $g$ is the string
coupling (2) $B^{(1)}$ is replaced by $B^{(2)}$ and (3) the dilaton
$\phi$ is replaced by $-\phi$. The description of the underlying
string states is however rather different and leads to a rather
interesting picture for the entropy.

The low lying states of the D-string are described by the lowest
modes of the Dirichelt open string theory. The physical single string
states of lowest (zero) mass are given by eight transverse vectors
and their supersymmetric partners. These can move only along the
D-string due to Dirichlet boundary conditions and being massless can
be either left or right moving along the D-string. Now consider many
such open strings moving along the D-string. Then a collection of
such strings all moving in the same direction (i.e. all left or all right
handed) constitues a BPS saturated state with some net momentum along
the string. These are the dual analogs of the NS-NS charged states
described by (\ref{eq:nine}).
Note that the length of a D-string wound $n_w$ times around
the $X^9$ direction is $n_w L$ so that the momenta of the individual
open string states can be ${2\pi m \over n_w L}$
for integer $m$. However we require
the {\em total} momentum of the state to be ${2 \pi n_p \over L}$.
We thus have
\ben
n_p = \sum_{i} \sum_m n_m^{(i)}{m \over n_w}
\label{eq:eleven}
\een
where $n_m^{(i)}$ denotes the number of single open string states with
momentum ${2 \pi m \over n_w L}$ and $(i)$ is the label for the vector
or spinor index of the state. 
We immediately see that the counting is exactly the same as for the
NS-NS states since there are eight transverse directions for the
D-string. This is of course expected from duality. The crucial difference
between the elementary string and the D-string descriptions is that
in the former a large number of excitations of {\em single string}
modes was responsible for the entropy. For the D-string,
however, the excitations are {\em multiple string} states of the Dirichelt
open strings.

The BPS states described above correspond to extremal black holes with
vanishing horizon. It was proposed in \cite{larswilc} that the same
philosophy should be valid for extremal holes with a nonzero horizon
area \cite{cvetic}. Recently string theoretic understandings of
such solutions have been achieved in terms of D-branes
\cite{stromvafa} \cite{callmal} and
the result is that the counting problem leads to precisely to the
formula (\ref{eq:one}).

\subsection{Non-BPS states : Hawking radiation ?}

The BPS black holes and black strings described above are stable or
marginally stable objects and do not radiate. The interesting objects
to consider are non-BPS states.  Such a state is unstable and will
decay to a stable BPS extremal state, and the real question is : does
this decay resemble Hawking radiation ?  

Let us discuss these non-BPS states in the D-string spectrum
\cite{dasmath}. For large
$L$ the relevant excitations are again multiple open string states but
now the stuck strings can move in either direction along the D-string.
We may excite a non-BPS state starting with a BPS state characterized
by $n_p$ by adding pairs of stuck open strings moving in opposite
directions so that the total momentum is still $2\pi n_p /L$. If we
add $n$ such pairs the entropy obtained from counting is
\ben
S_{non-BPS} = 2\pi [{\sqrt{2(|n_p| + n)n_w}} +{\sqrt{2n n_w}}]
\label{eq:twelve}
\een
In perturbation theory the mass of such a state is given by
\ben
M = M_{BPS} + {4\pi n \over L}
\label{eq:thirteen}
\een
which agrees with the mass of corresponding objects in the elementary
string spectrum upto terms of order $O(1/TL^3)$. Furthermore the 
inelastic threshold for excitation of such a state, ${4 \pi \over L}$ agrees
with the corresponding threshold for excitation of a NS-NS charged
state as calculated in \cite{mandalwadia}.
For finite $L$ this formula 
is valid for weak coupling, but for large $L$ corrections to the mass
formula are suppressed by terms of order $O(g/Tn^2L^3)$ and one can trust
the masses and the counting for moderately strong coupling as well. 
A pair of such open stuck strings may annahilate and decay into a
closed string which may escape out to infinity.  Note that the decay
rate is also suppressed by inverse powers of $L$
\cite{dasmath} \cite{dasmatha}
. Thus for large $L$ the states are almost stable and 
the counting of states makes sense. Non-BPS states have been
also discussed in \cite{callmal} and \cite{stromhor} for cases
where the extremal solution has nonzero horizon area and recently
in \cite{klebah} for extremal solutions which have a vanishing
horizon area.

The above scenario for black string entropy makes sense if the decay
of such a non-BPS state resembles Hawking radiation. The reason why it
might be like radiation is that there are a large number of states
for a given  value of the macroscopic parameters (mass and charge) and
one should sum over the initial states. This same degeneracy of states 
is responsible for the entropy (\ref{eq:twelve}). Thus outgoing particles
would have a thermal distribution with a temperature $T = ({\partial S
\over \partial E})^{-1}$ where $E$ is the energy of the outgoing
particle. Noting that $E = {4\pi n \over L}$ in this decay process
and using (\ref{eq:twelve}) one gets a ``temperature'' for $n << n_p$ 
\ben
T = {2 \over L} {\sqrt{2n}}
\label{eq:fourteen}
\een

Is this the Hawking temperature ? A slightly different but equivalent
calculation in \cite{callmal} for decay of non-BPS excitations above
BPS black holes with nonzero horizon area (obtained by compactifying
the 1-brane together with another 5-brane) showed that there is an
exact agreement between the temperature calculated in this fashion and
the Hawking temperature of the non-extremal black hole. However, this
calculation is valid for weak coupling where the horizon is actually
smaller than the string scale (see also \cite{stromhor}). We believe
that the large $L$ limit may be more tractable for reasons given above.

\subsection{Space-time interpretation}

For black strings carrying NS-NS charges the
Beckenstein Hawking entropy came from the large number of oscillator
degrees of freedom of the string state. So far as the macroscopic
object is concerned these are ``internal'' degrees of freedom.

For D-strings however something remarkable has happened. The entropy
comes from the physical motion of the lowest modes of the stuck open strings 
along the D-string. These are in fact momentum modes of a {\em
field theory} of massless particles (eight scalars and eight fermions)
on the D-brane worldvolume - the
supersymmetric gauge theory discussed in \cite{witten}. These degrees
of freedom on the brane volume give rise to the microstates necessary
for the thermodynamic behavior of the black string (or hole).

What he have here is almost normal thermodynamics of massless particles 
moving on the two dimensional worldsheet of the D-string. However since
there is a net momentum along the D-string the right and left moving
modes have different total energies. The left movers have a total
energy of $[2\pi (n_p + n)/L]$ while the right movers have energy
$[2\pi n /L]$. For $d$ massless scalars and $d$ massless fermions
standard thermodynamics predicts that the entropy is
\ben
s = {\sqrt{\pi d L E \over 2}}
\label{eq:fifteen}
\een
for left and right movers separately. With the appropriate expression
for the energies  and $d = 8$ for the eight transverse directions
this reduces to the two terms in (\ref{eq:twelve}). These particles
have a ``temperature'' which is the temperature of the equivalent
canonical ensemble, this is $\theta = {\sqrt{8 E \over \pi d L}}$, i.e.
\bea
\theta &=& {1 \over L}{\sqrt{2n}}~~~~~~~~{\rm for~ right~ movers} 
\nonumber \\
\theta &=& {1 \over L}{\sqrt{2(|n_p| + n)}}~~~~~~~~
{\rm for~ left~ movers} 
\label{eq:sixteen}
\eea
Note that in defining these temperatures we have kept the charges
$n_p$ and $n_w$ fixed.
Furthermore these temperatures are not the same as the candidate Hawking
temperature in (\ref{eq:fourteen}). A similar temperature of left
movers have been defined in \cite{callmal}, but differ by a factor
of two from the above.

The D-string description thus gives a useful physical picture of black
hole thermodynamics. There are these open strings stuck to the
D-string. The outside observer in the asymptotic region cannot see the
details of the motion of these strings along the brane and therefore
decides to average over them : this gives rise to thermodynamics in a
way entirely similar to the way thermodynamics appears in ordinary
processes. It is likely that this rather simple physical picture will
be useful to study the black hole problem.

Note that for the D-string the target space fields of the open string
become coordinates on the D-string worldsheet. Duality of the IIB
theory ensures that the oscillator count of the elementary string is
the same as the count of open string momentum modes. However for
D-branes of higher dimensionality the open string massless fields are
coordinates on a {\em higher} dimensional worldvolume. Nevertheless
the thermodynamics is that of some physical particles moving on this
worldvolume \footnote{For an example on 3-branes see \cite{klebah}}.

\section{The Horizon as a D-brane}

In this section we briefly mention some speculative ideas about
entanglement entropy in string theory and its relation to
D-branes \cite{dasmathb}.  As discussed
in \cite{suss1} and \cite{suss2} in string theory it is tricky to
define an entanglement entropy since a string can be partly inside and
partly outside the horizon. To an outside observer using coordinates
which stop at the horizon, such strings would actually appear as {\em
open strings} whose ends move along the horizon. But this means that
the horizon behaves like a D-brane !

To make this idea a little more concrete we can imagine working in a
fixed gauge. Consider closed string theory in flat space with an
imaginary divide at $x = 0$, like in Section 1. The wave functional of
a string state may be schematically written as
\ben
\Psi [ \Phi(X_L (\sigma)), \Phi(X_R (\sigma)), \Phi(X_H (\sigma)); t]
\label{eq:seventeen}
\een
where $X_L (X_R)$ denote strings which are entirely in $x < 0$ ($x >
0$) and $X_H$ denote strings which are half inside and half outside
and $\Phi$ the corresponding string fields. Roughly speaking the
string field $\Phi_H$ may be expressed as a product of open string
fields living on the left and the right. Then we may proceed to define
an entanglement entropy by integrating over the closed string fields on
the left as well as the open string fields on the left. The density
matrix is now functionals of a closed string field $\Phi_R$ and an
open string field $\Phi_{O,R}$ and the latter corresponds to open
strings which are stuck on the horizon. A fixed area path integral
would now involve both closed and open string diagrams.
For a black hole one would have the euclidean black hole
background with a conical singularity. It is likely that such quantities 
may be obtained by using D-brane technology. We note while the
above picture is related to (and inspired by) \cite{suss1} \cite{suss2}
there may be some differences.

In fact the above picture immediately leads to a finite contribution
to the black hole entropy proportional to the horizon area from the
open string states stuck to the horizon. This is because the effective
action for the modes of such strings is in fact a Nambu-Goto action on
the D-brane worldsheet, coming from the disc diagram. It is important
to note that we are doing an euclidean calculation so that, e.g. in
four dimensions, the space has a topology of $S^2 \times R^2$ and the
horizon is actually the $S^2$ at the origin of $R^2$ (and not $S^2 \times
R$). 

This picture of viewing the horizon itself as a D-brane could give
a realization of the membrane paradigm in string theory.
Furthermore this picture looks tantalizingly similar to the more
concrete picture of black string thermodynamics in the previous
section if the D-string is actually the horizon rather than the
singularity. In fact in \cite{callmal} it was found that the classical
geometry for the 1-brane and 5-brane configurations seem to show
that these branes are actually located at the horizon. While we
do not fully understand how and why this happens, it is possible that
this fact will help us to concretize the picture of the horizon as
a D-brane.

\section{Acknowledgements}

It was a unique experience to be in Ho Chi Minh city. I would like to
express my sincere thanks to the organizers, particularly Tranh Van,
for inviting me to this wonderful conference.  I would like to thank
Samir Mathur for collaboration on the work described in sections 2 and
3, for numerous discussions, and comments on the manuscript.  I would
also like to thank N. Ishibashi, A. Jevicki, H. Kawai, A. Sen and S.
Wadia for useful discussions.  Finally I thank the Theory Group at KEK
for hospitality.

\end{document}